\documentclass[amsmath,amssymb,aps]{revtex4}

\selectfont 

\usepackage{graphicx}       % Include figure files
\usepackage{dcolumn}        % Align table columns on decimal point
\usepackage{bm}             % bold math
\usepackage{psfrag}

\newcommand{\rco}{\hat{r}_{\circ}}
\newcommand{\rcpm}{\hat{r}_{\pm}}
\newcommand{\rcp}{\hat{r}_{+}}
\newcommand{\rcm}{\hat{r}_{-}}
\newcommand{\bco}{{b}_{\circ}}
\newcommand{\dco}{{d}_{\circ}}

\newcommand{\beq}{\begin{equation}}
\newcommand{\eeq}{\end{equation}}

\begin{document}

\preprint{}

 \title{The Quasinormal Mode Spectrum of a Kerr Black Hole in the Eikonal Limit}
 
\author{Sam R. Dolan}
 \email{s.dolan@soton.ac.uk}
 \affiliation{%
 School of Mathematics, University of Southampton, Highfield, Southampton SO17 1BJ, UK.
}%

\date{\today}

\begin{abstract}

It is well established that the response of a black hole to a generic perturbation is characterized by a spectrum of damped resonances, called quasinormal modes; and that, in the limit of large angular momentum ($l \gg 1$), the quasinormal mode frequency spectrum is related to the properties of unstable null orbits. In this paper we develop an expansion method to explore the link. We obtain new closed-form approximations for the lightly-damped part of the spectrum in the large-$l$ regime. We confirm that, at leading order in $l$, the resonance frequency is linked to the orbital frequency, and the resonance damping to the Lyapunov exponent, of the relevant null orbit. We go somewhat further than previous studies to establish (i) a spin-dependent correction to the frequency at order $1 / l$ for equatorial ($m = \pm l$) modes, and (ii) a new result for polar modes ($m= 0$). We validate the approach by testing the closed-form approximations against frequencies obtained numerically with Leaver's method. 

\end{abstract}

\maketitle

% Paper Plan:
%
% Introduction:
%   

\section{Introduction}

Quasinormal modes (QNMs) are damped resonances which play a key role in black hole dynamics \cite{Kokkotas-Schmidt, Nollert, Ferrari-Gualtieri, Berti-Cardoso-Starinets}. For example, after the merger of a pair of black holes, the composite system `rings down' through gravitational wave emission until it settles down into a quiescent axisymmetric Kerr phase. The gravitational wave signal from ringdown is dominated by the least-damped QNM resonances. Chandrasekhar \cite{Chandrasekhar} was moved to compare the ringdown signal to the dying pure notes sounded by a bell. The analogy is apt; when a bell is struck, the dying tones depend only on the properties of the bell, rather than the hammer used to strike it, which affects only the relative degree of excitation of overtones. Likewise, the QNM spectrum depends only on the underlying properties of the black hole, rather than the complicated details of any initial perturbation.

Physically, a black hole quasinormal mode is a decaying resonance which satisfies a pair of causally-motivated boundary conditions (typically being ingoing at the event horizon, and outgoing at spatial infinity). Mathematically, `quasinormal ringing' emerges from a sum of residues of poles of the Green function in the complex frequency domain \cite{Leaver-1986}. Each pole corresponds to a quasinormal mode of a single complex frequency.  The real part of the frequency corresponds to the oscillation rate and the (negative) imaginary part corresponds to the damping rate. 

In the Kerr spacetime QNMs are labelled by three indices: multipole $l$, azimuthal number $m$ and overtone number $n \ge 0$. The QNM spectrum depends only on the properties of the field (i.e. spin $s$) and the underlying black hole geometry (i.e. mass $M$, charge $Q$, angular momentum $J=aM$). In this paper, we concern ourselves with the most astrophysically-relevant case: the uncharged non-extremal Kerr black hole ($0 \le J < M^2$ and $Q=0$).

A wide range of methods have been developed for determining QNMs. A non-exhaustive list \cite{Berti} includes (i) time domain methods \cite{Vishveshwara, Dorband};
(ii) direct integration in the frequency domain \cite{Chandrasekhar-Detweiler}; 
(iii) inverse potential methods \cite{Blome-Mashhoon, Ferrari-Mashhoon}; 
(iv) WKB methods \cite{Mashhoon-1983, Schutz-Will-1985, Iyer-Will, Iyer, Seidel-Iyer, Kokkotas, Konoplya};
(v) phase-integral methods \cite{Phase-integral1, Phase-integral2, Natario-Schiappa};
(vi) continued fraction method \cite{Leaver-1985, Nollert-1993}; 
(vii) a semi-classical expansion method \cite{Dolan-Ottewill}; and
(viii) an asymptotic iteration method \cite{Cho-Cornell-Doukas-Naylor-2009, Cho-Cornell-Doukas-Naylor-2010}.
Review articles \cite{Kokkotas-Schmidt, Nollert, Ferrari-Gualtieri, Berti-Cardoso-Starinets} explore some of these methods in greater depth.

The quasinormal frequencies of the Kerr black hole were first computed by Detweiler \cite{Detweiler}, and later by Leaver \cite{Leaver-1985}, Seidel and Iyer \cite{Seidel-Iyer}, Kokkotas \cite{Kokkotas} and Onozawa \cite{Onozawa} (among others). In essence, the problem of accurately determining the QNMs of the Kerr black hole numerically has been `solved' for a quarter of a century; the method introduced by Leaver \cite{Leaver-1985} is fast, reliable and accurate, and with certain modifications \cite{Nollert-1993}, robust at large overtones $n$ and angular momenta $l, m$. 
However, a numerical method provides little by way of physical insight. In contrast, approximation methods aim to provide some insight, perhaps at the expense of accuracy and applicability.  For example, a body of work on the WKB method \cite{Mashhoon-1983, Schutz-Will-1985, Iyer-Will, Iyer, Seidel-Iyer, Konoplya} in spherically-symmetric spacetimes `explains' the relationship between the low-overtone frequencies and the shape of the peak of the radial potential \cite{Iyer-Will}. The low-overtone QNM spectrum of the Kerr black hole can be obtained via the WKB method \cite{Seidel-Iyer, Kokkotas}, although the expansion formulae are somewhat complicated. % and in their raw form provide relatively little geometric insight. 

It has been known for many years that quasinormal modes are intimately linked to the existence and properties of unstable photon orbits \cite{Goebel, Ferrari-Mashhoon}. A quarter-century ago, Mashhoon \cite{Mashhoon} examined an aggregate of test null rays in orbit in the equatorial plane of a Kerr-Newman ($Q > 0$) black hole near the co-rotating circular orbit. Mashhoon's analysis suggested that, in the eikonal regime ($l \gg 1$), the maximally-corotating mode ($m=l$) has the frequency 
\beq
\omega_{ln}^{m=l} \approx l \omega_+ - i \lambda_+ (n+1/2) \label{mashhoon-eq}
\eeq
where $\omega_+$ is the Kepler frequency for null rays on the unstable orbit and $\lambda_+$ is the decay rate of the unstable orbit (subsequently identified as the Lyapunov exponent \cite{Lyapunov1, Lyapunov2, Lyapunov3}). These orbital quantities were plotted as a function of $a$ in \cite{Ferrari-Mashhoon-letter}; see also our Fig.~\ref{fig-orbparams}. The `geometric-optics' approach of Mashhoon is both appealing and insightful. For example, it neatly `explains' why the maximally corotating modes ($m = l$) are faster-oscillation and slower-decaying than the other modes ($m < l$). However, Mashoon's original analysis was only `indirect', in the sense that the Teukolsky equation, which describes massless perturbations of the Kerr spacetime, was not solved directly. So, for instance, it is not clear from (\ref{mashhoon-eq}) what effect the spin of the perturbing field has on the frequency spectrum.

A recent work by Hod \cite{Hod} demonstrated that a bridge may be built between ``geometric-optics'' intuition and ``Teukolsky equation'' exactitude.  By analysizing the Teukolsky equation in the extremal limit ($a \rightarrow M$), Hod showed that (\ref{mashhoon-eq}) is indeed good approximation.  %This is an important regime, since authors have found that the damping of the fundamental mode goes to zero in this limit. 

The aim of this paper to show that improved approximations to QNM frequencies may be obtained by directly analysing the Teukolsky equation, without recourse to WKB methods. We go further than other studies (such as Mashhoon \cite{Mashhoon} and Hod \cite{Hod}) to obtain frequency expansions in closed form which include the leading-order effect of the spin of the perturbing field, and which are valid outside the extremal regime (i.e. for all $a$). As well as considering the equatorial modes ($m = \pm l$) we also analyse the `polar' ($m=0$) modes (see Fig.~\ref{fig-geodesics} for illustration). We  confirm that the expansions provide the correct fits to the numerically-determined spectrum, in the limit $l \gg 1$.

% More generally, it was recently demonstrated \cite{Lyapunov1} that, in any static spherically symmetric asymptotically flat spacetime, to leading order in $l$ the QNM frequency is $
%\lim_{l \rightarrow \infty} \text{Re} ( \omega_{ln} ) =  \Omega l , \quad \quad \quad \quad \lim_{l \rightarrow \infty} \text{Im} ( \omega_{ln} )  = - (n + 1/2) |\lambda| $
%where $\lambda$ is the so-called \emph{Lyapunov exponent}: the (inverse of the) instability timescale of the unstable orbit \cite{Lyapunov2, Lyapunov3}. 

To obtain our new analytic results, we apply a recently-introduced expansion method \cite{Dolan-Ottewill}. Previously, this method has been applied to a range of spherically-symmetric spacetimes \cite{Dolan-Ottewill}, and to a simple cylindrically-symmetric system (the `draining vortex') \cite{Leandro}. In this paper, we extend the method to treat an axisymmetric system for the first time. The key idea behind our approach is that the properties of modes of high angular momentum $l \gg 1$ are related to the properties of the unstable photon orbits of the spacetime (see Fig.~\ref{fig-geodesics} and \cite{Teo}). This idea is not new \cite{Goebel}; the novel ingredient is a `semi-classical' ansatz enables us build a bridge between the classical `geodesic' picture of a battery of null geodesics near the photon orbit \cite{Mashhoon}, and a perturbative analysis at the level of the Teukolsky equation. This ansatz allows the QNM frequencies and wavefunction to be expanded in inverse powers of $l$. 

The remainder of this paper is organized as follows: in Sec.~\ref{sec:geodesics} we describe the Kerr spacetime, null geodesics, and the Teukolsky equation; in Sec.~\ref{sec:expansion-method} we develop the expansion method and obtain the key analytic results; in Sec.~\ref{sec-validation} we validate the approach by testing the expansion against numerical results. We conclude in Sec.~\ref{sec:conclusion} with a brief discussion.

\section{Spacetime, Geodesics and Waves\label{sec:geodesics}}
In this section we lay some groundwork, and give geometrical motivation for the QNM analysis which follows in later sections. In Sec.~\ref{subsec:spacetime} we review some properties of the Kerr spacetime; in Sec.~\ref{subsec:geodesics} we consider the null geodesics of the spacetime, and the unstable photon orbits \cite{Teo}; and in Sec.~\ref{subsec:perturbations} we recap the theory of weak-field perturbations of Kerr.

\subsection{The Kerr Spacetime\label{subsec:spacetime}}

\begin{figure}
 \includegraphics[width=7cm]{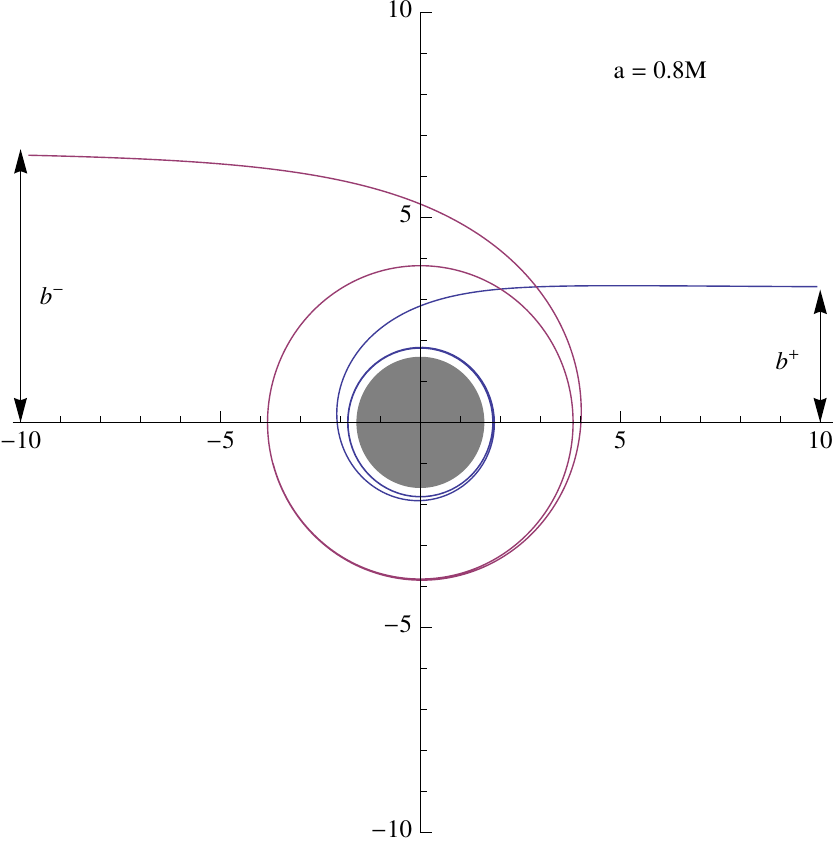}
  \includegraphics[height=8cm]{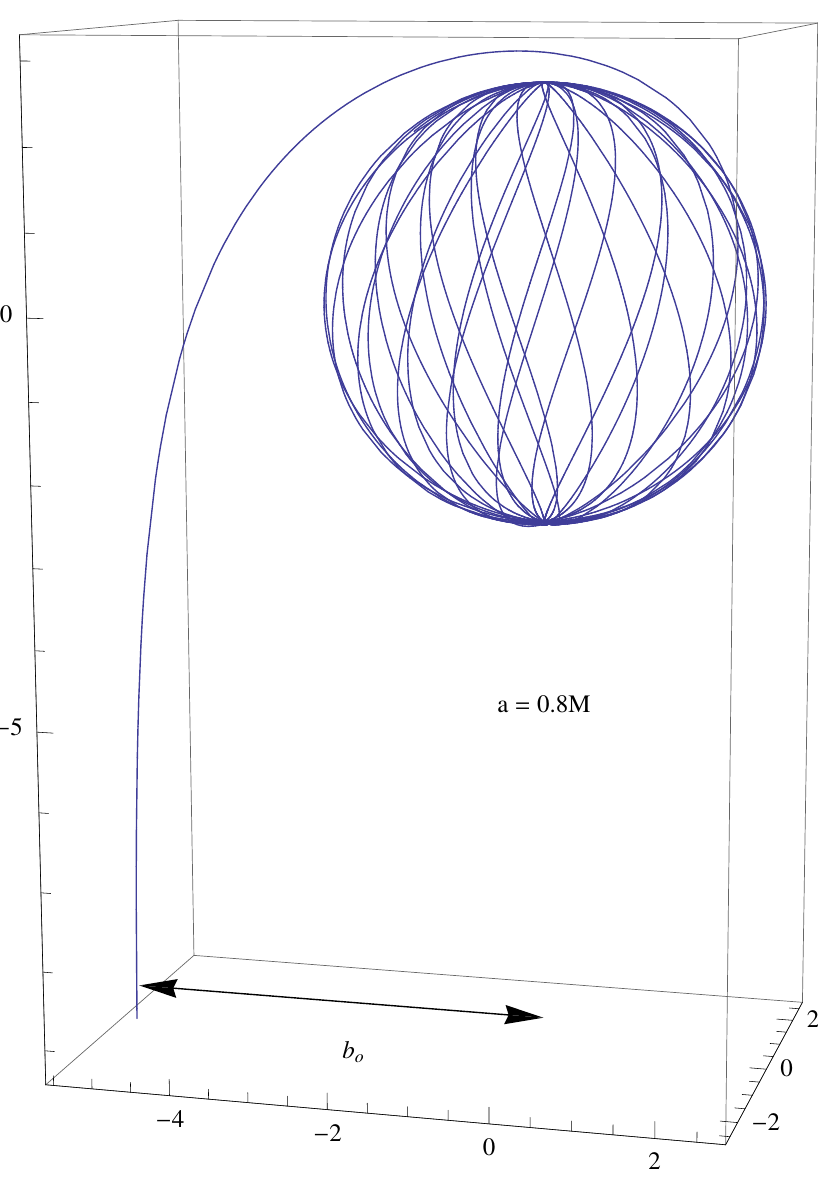}
 \caption{\emph{Critical null geodesics for $a = 0.8M$}. These plots show the special null geodesics which impinge from infinity and asymptote to the unstable photon orbits. The 2D plot (left) shows co- and counter-rotating orbits in the equatorial plane (with the spin axis pointing out of the page). The 3D plot (right) shows a polar orbit, with the spin axis of the BH pointing up the page.}
 \label{fig-geodesics}
\end{figure}

\begin{figure}
 \includegraphics[width=7.5cm]{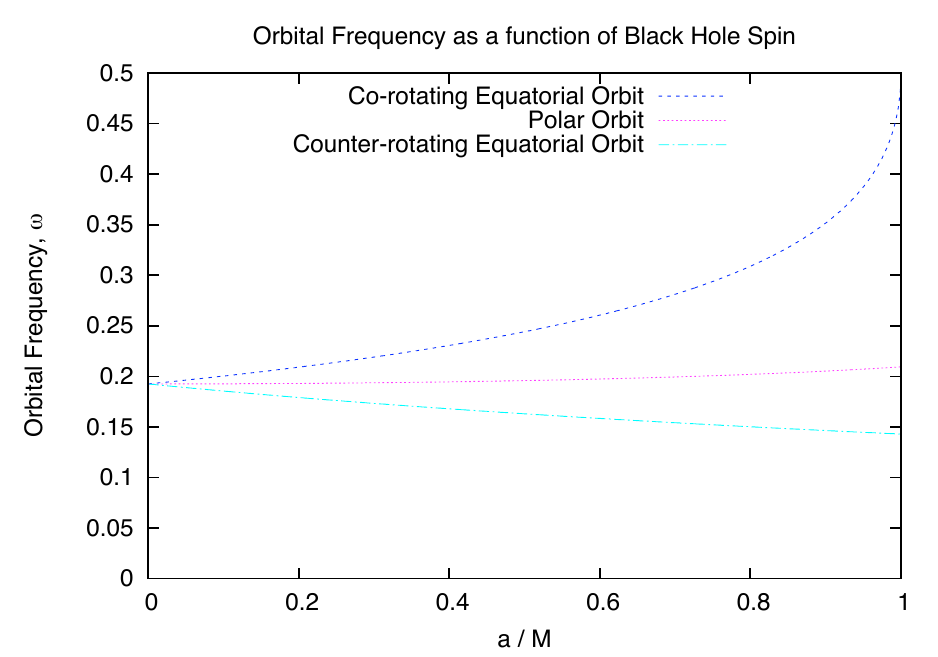}
  \includegraphics[width=7.5cm]{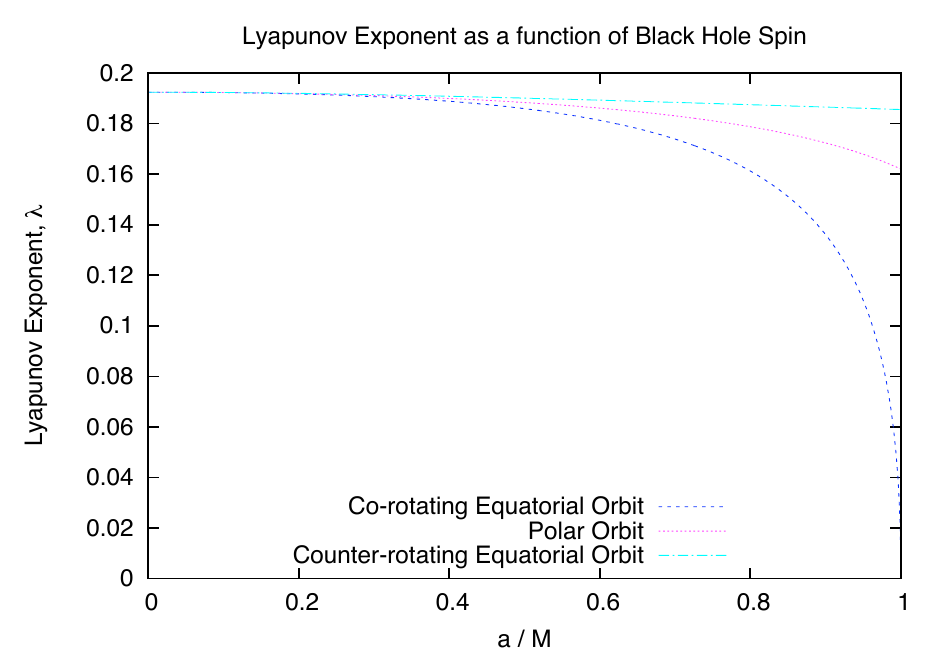}
 \caption{\emph{Parameters for Unstable Photon Orbits around a Kerr black hole as a function of $a = J / M$}. The left plot shows the orbital frequency with respect to coordinate time $t$ (time measured at spatial infinity). The right plot shows the Lyapunov exponent, which controls the decay rate of QNMs. The lines are for the special cases of prograde and retrograde equatorial orbits (see also Fig.~1 \& 2 in \cite{Ferrari-Mashhoon-letter}) and the polar orbit, discussed in Sec.~\ref{subsec:geodesics}. General orbits would correspond to lines between these limits. }
 \label{fig-orbparams}
\end{figure}

The Kerr metric \cite{Kerr-1963} describes the spacetime around a rotating uncharged black hole in vacuum (for details see e.g. \cite{Chandrasekhar, MTW, WiltshireVisserScott}). In the Boyer-Lindquist coordinate system $\{t,r,\theta,\phi\}$ the line element takes the form
\beq
ds^2 = -(1 - 2Mr/\rho^2) dt^2 - (4aMr \sin^2 \theta / \rho^2) dt d\phi + (\rho^2/\Delta) dr^2 + \rho^2 d\theta^2 + \sin^2\theta(r^2 + a^2 + 2Ma^2 r \sin^2\theta/\rho^2 ) d \phi^2,
\eeq
where $\Delta \equiv r^2 - 2Mr + a^2$ and $\rho^2 \equiv r^2 + a^2 \cos^2 \theta$, and where $M$ and $aM$ are the mass and angular momentum of the black hole. The spacetime has an ergosphere within $r < 2M$ and horizons at $r = r_{h\pm} \equiv M \pm \sqrt{M^2-a^2}$. % More detail can be found in a range of texts \cite{MTW,

\subsection{Null Geodesics and Orbits\label{subsec:geodesics}}
%Light rays follow null geodesics. %, which may be deduced from the Euler-Lagrange equations for the Lagrangian $\mathcal{L} = \tfrac{1}{2} g_{\mu\nu} \dot{x}^\mu \dot{x}^\nu$, together with the null condition $\mathcal{L} = 0$ (where $\dot{}$ denotes differentiation wrt an affine parameter). 
Killing vector fields (satisfying $\xi_{(a;b)} = 0$) associated with time translation $\xi^a_{(t)}$ and rotation around the symmetry axis $\xi^a_{(\phi)}$ give rise to two constants of motion along a geodesic with tangent vector $u^a$: an `energy' $E \equiv -\xi^a_{(t)} u_a$ and an `azimuthal angular momentum' $L_z \equiv \xi^a_{(\phi)} u_a$. The existence of a Killing tensor $K_{ab}$ satisfying $K_{(ab;c)} = 0$ implies a further constant of motion, the `Carter constant'  \cite{Carter-1968} defined as $\mathcal{Q} = K_{ab} u^{a} u^{b} + L_z^2 - E^2$. The Carter constant governs the motion of geodesics in the latitudinal direction. The equations of motion for null rays can be written in first order form \cite{Chandrasekhar, Teo, MTW, Drasco-Hughes-2004} as
\begin{eqnarray}
\rho^2 \dot{t} &=& \left( \Sigma^2 E - 2Ma r L_z \right) / \Delta ,  \label{t-eq} \\
\rho^2 \dot{\phi} &=& \left( 2MarE + (\rho^2 - 2Mr) \frac{L_z}{\sin^2 \theta} \right) / \Delta \label{phi-eq} \\
\left( \rho^2 \dot{r} \right)^2 &=& E^2 r^4 + (a^2 E^2 - L_z^2 - \mathcal{Q}) r^2 + 2M\left[(aE - L_z)^2 + \mathcal{Q} \right] r - a^2 \mathcal{Q} ,  \label{r-eq} \\
\left( \rho^2 \dot{\theta} \right)^2 &=& \mathcal{Q} - \left[ \frac{L_z^2}{\sin^2 \theta} - a^2 E^2 \right] \cos^2 \theta , \label{theta-eq}
\end{eqnarray}
where $\Sigma^2 = (r^2 + a^2)^2 - a^2 \Delta \sin^2 \theta$ and an overdot denotes differentiation with respect to an affine parameter. Note that the $r$ and $\theta$ equations may be decoupled into ordinary differential equations by using the `Mino time' \cite{Mino, Drasco-Hughes-2004} parameter $\nu$ defined by  $d \nu = d \lambda / \rho^2$. For our purposes this is not necessary.

In the following section we consider three sets of constant-$r$ null orbits: % (which are unstable but nevertheless perpetual if initial conditions are perfectly tuned): 
\emph{prograde} and \emph{retrograde equatorial} orbits with $\mathcal{Q} = 0$ and \emph{polar} orbits with $L_z =0 $. These orbits are illustrated in Fig.~\ref{fig-geodesics}.  The cases are henceforth distinguished using subscripts $+$, $-$ and $\circ$, respectively; for example, we will denote the orbital radii by $\rcp$, $\rcm$ and $\rco$ and the angular frequencies (with respect to coordinate time $t$) by $\omega_+$, $\omega_-$ and $\omega_\circ$. Figure \ref{fig-orbparams} shows the orbital frequency and Lyapunov exponent for these special orbits as a function of black hole rotation $a$.

\subsubsection{Equatorial orbits}
Consider first null geodesics in the equatorial plane, with $\mathcal{Q} = 0$. The ratio of azimuthal angular momentum to energy defines an \emph{impact parameter} (see Fig.~\ref{fig-geodesics}),
\beq
b = L_z / E .
\eeq
%which is the perpendicular distance to the black hole centre for a null geodesic approaching from infinity in the equatorial plane. 
Null geodesics in the equatorial plane are governed by the orbital equation
\beq
\dot{r}^2 = \frac{L_z^2}{b^2} \left[ 1 + \frac{a^2 - b^2}{r^2} + \frac{2M(b-a)^2}{r^3}  \right]  \label{equatorial-orbit-eq}
\eeq
Perpetual photon orbits are possible if conditions $\dot{r} = 0$ and $\ddot{r} = 0$ can be satisfied simultaneously. It is straightforward to show that both prograde $(+)$ and retrograde $(-)$ orbits exist, at orbital radii $\rcpm$, 
\beq
\rcpm = 2M \left[ 1 + \cos\left(\tfrac{2}{3} \cos^{-1}(\mp a / M) \right)  \right], 
\eeq
with critical impact parameters $b_\pm$ given by
\beq
b_\pm = \pm 3\sqrt{M \rcpm} - a  .  \label{bcrit}
\eeq
Note that we define the critical impact parameter $b$ so that it is negative for the retrograde trajectory. 
The orbital frequency $\omega_\pm = \dot{\phi} /  \dot{t} $ of the circular orbit is obtained from the ratio of the $\phi$ and $t$ equations (\ref{t-eq} and \ref{phi-eq}), 
and it is straightforward to show that it is simply
\beq
\omega_{\pm} = \frac{1}{\left| b_\pm \right|}   = \frac{M^{1/2}}{(\rcpm)^{3/2} \pm a M^{1/2}}  .  \label{kepler-freq}
\eeq
In the critical case, the orbital equation (\ref{equatorial-orbit-eq}) can be factorized into
\beq
\dot{r}^2 = \frac{L_z^2}{b_\pm^2} \left(1 - \frac{\rcpm}{r} \right)^2 \left(1 + \frac{2 \rcpm}{r} \right)  .\label{equatorial-orbit-crit}
\eeq

In \cite{Mashhoon, Lyapunov1, Hod} it was established that the decay rate for the equatorial circular orbits is controlled by Lyapunov exponent $\lambda$ (see Eq.~(\ref{mashhoon-eq})). For the equatorial orbits \cite{Lyapunov1}, it has been shown that the Lypanov exponent is 
\beq
\lambda_{\pm} = \frac{1 - 2x_\pm}{\left|b_\pm \right| \sqrt{1 - x_\pm^2}}   \label{lyapunov-eq}
\eeq
where $x_\pm$ is the ratio
\beq
x_\pm \equiv a / b_\pm
\eeq
Note that $b_- < 0$ and hence $x_- < 0$. 

In the Schwarzschild limit ($a \rightarrow 0$) we recover $\hat{r}_\pm = 3M$,  $b_\pm = \sqrt{27} \, M$ and $\omega_\pm = \lambda_\pm = 1/(\sqrt{27} M)$. In the extremal limit ($a \rightarrow M$) we have $\rcm = 4M$, $b_- = -7M  \Rightarrow \omega_- = 1/(7M)$, $\lambda_- = 3/(7\sqrt{5}M)$ and $\rcp = M$, $b_+ = 2M \Rightarrow \omega_+ = 1/(2M)$ and $\lambda_+ = 0$. Note that the decay rate of the co-rotating mode tends to zero in the extremal limit (as many authors have previously observed).

%An alternative derivation of the Lypaunov exponent (\ref{lyapunov-eq}) is given in Appendix ...

\subsubsection{Polar orbits\label{subsec-polar-geodesics}}
Consider next the special case of trajectories with zero azimuthal angular momentum ($L_z = 0$), which we will refer to as `polar' geodesics. For polar geodesics we may define the natural impact parameter
\beq
b^2 = \mathcal{Q} / E^2 + a^2
\eeq
which corresponds to the perpendicular distance between the geodesic and the rotation axis (measured at spatial infinity). Note that polar geodesics are parallel to the rotation axis at spatial infinity, as shown in Fig.~\ref{fig-geodesics}.

Null polar geodesics are governed by the equation
\beq
\left( E^{-1} \rho^2 \dot{r} \right)^2 \equiv R(r) = r^4 + (2a^2 - b^2) r^2 + 2 M b^2 r + a^2 (a^2 - b^2)  \label{R-defn}
\eeq
An unstable orbit arises at radius $r=\rco$ where $R(\rco) = 0$ and $R^\prime(\rco) = 0$, given by
\beq
\rco = M + 2 \sqrt{M^2 - a^2/3} \, \cos \left[ \frac{1}{3} \cos^{-1} \left( \frac{M(M^2-a^2)}{(M^2 - a^2/3)^{3/2}} \right) \right] 
\eeq
(see e.g. Eq.~(14) in \cite{Teo}). The critical impact parameter is given by
\beq
\bco^2  %= a^2 + \rco^2 \left( \frac{3 \rco^2 + a^2}{\rco^2 - a^2} \right) 
=  \frac{(3 \rco^2 - a^2)(\rco^2 + a^2)}{\rco^2 - a^2} ,   \label{bcrit-polar}
\eeq
and in the critical case $b = b_\circ$, the radial equation (\ref{R-defn}) can be factorized into
\beq
R(r) = \left( r-\rco \right)^2 \left( r^2 + 2 \rco r - a^2 (\bco^2 - a^2) / \rco^2 \right) .  \label{R-factorization}
\eeq
The period $T_\circ$ for a null geodesic on the unstable polar orbit at $r=\rco$ can be found from integrating the ratio of the $t$ and $z = \cos(\theta)$ equations (see Appendix \ref{appendix:polar} for details),
%\begin{eqnarray}
%\left( \rho^2/E \right)^2 \dot{t}^2 &=& \left( \Sigma^2 / \Delta \right)^2 \\
%\left( \rho^2/E \right)^2 \dot{z}^2 &=& \bco^2-a^2 - (\bco^2-2a^2)z^2 - a^2 z^4 
%\end{eqnarray}
%to obtain
\beq
T_\circ = 2 \int_{-1}^{+1} \frac{d t}{d z} dz = 2 \dco  \int_{-1}^{+1} \left( \frac{1+ a^2 z^2 / \dco^2}{1 - z^2} \right)^{1/2}  dz = 4 \dco \, \text{ellipE}(i a / \dco) .   \label{polar-period}
\eeq
where we have defined $\dco = \sqrt{\bco^2 - a^2} \equiv \sqrt{Q}/E$, and $\text{ellipE}(\cdot)$ is the complete elliptic integral of the second kind \cite{Abramowitz-Stegun} (a note of caution: $\text{ellipE}(k)$ is implemented in Maple as $\text{EllipticE}(k)$, but in Mathematica as $\text{EllipticE}(k^2)$). 
Hence the orbital frequency is 
\beq
\omega_\circ = \frac{\pi }{2 \dco \, \text{ellipE}(i a / \dco)} .   \label{polar-orbfreq}
\eeq

We expect the imaginary part of the QNM of the polar modes ($L_z = 0$) to be related to the Lyapunov exponent $\lambda$ for the polar null orbits. In Appendix \ref{appendix:polar} it is shown to be
\beq
\lambda_\circ = \frac{\rco}{(\dco)^2} \frac{\text{ellipK} \left( x_\circ / \sqrt{1+x_\circ^2} \right)} {\sqrt{1+x_\circ^2} \, \text{ellipE}( i x_\circ)} \left( 3 - \frac{a^2 \dco^2}{\rco^4}  \right)^{1/2} , \label{polar-lyapunov}
\eeq
where $\text{ellipK}(\cdot)$ is the complete elliptic integral of the first kind and we have defined the ratio
\beq
x_\circ = a / \dco. 
\eeq

In the extremal limit, $a \rightarrow M$, we have $\rco/M = 1+\sqrt{2} \approx 2.4142$ and $b_\circ / M \approx 4.8284$ and $M \omega_\circ \approx 0.20937$ and $M \lambda \approx 0.162006$. Unlike the corotating modes, the $m=0$ modes are still significantly damped in this limit (as is shown in Fig. \ref{fig-orbparams}).

\subsubsection{General orbits}
Above we considered two limiting cases: equatorial orbits restricted to $\theta = \pi / 2$, and polar orbits that explore the full range of polar angles, $0 \le \theta \le \pi$. A general orbit will explore a more limited range of polar angles, $\pi / 2 - \Delta \theta \le \theta \le \pi / 2 + \Delta \theta$. The angular range $\Delta \theta$ is a function of the ratio $L_z / \sqrt{\mathcal{Q}}$, with limiting cases $\Delta \theta( L_z / \sqrt{\mathcal{Q}} \rightarrow \infty) = 0$ and $\Delta \theta(L_z / \sqrt{\mathcal{Q}} \rightarrow 0) = \pi/2$. It was shown in e.g. \cite{Teo} that perpetual orbits exist for all ratios $L_z / \sqrt{\mathcal{Q}}$. A general analysis is left for future work.

\subsection{Perturbations of the Kerr Spacetime\label{subsec:perturbations}} 
To analyse gravitational dynamics around a rotating black hole, one may examine perturbations of the metric tensor via Einstein's equations linearized around the Kerr background (assuming the perturbations are ``small''). This approach yields a set of coupled partial differential equations, which are difficult to analyze. An alternative approach is to consider perturbations of the Weyl and Ricci scalars using the Newman-Penrose formalism. This was the approach taken by Teukolsky \cite{Teukolsky-1972, Teukolsky-1973}, who decoupled and separated the equations to reduce them to a single `master equation' for the radial part of the perturbation. 

A `master equation' describes massless perturbations of scalar ($|s|=0$), spinor ($|s|=1/2$), electromagnetic ($|s|=1$) and gravitational ($|s|=2$) types. In vacuum, the radial equation is
\beq
\Delta^{-s} \frac{d}{d r} \left( \Delta^{s+1} \frac{d R}{d r} \right) + \left( \frac{K^2 - 2is(r-M) K}{\Delta} + 4 i s \omega r - \Lambda \right) R = 0  \label{rad-eq-kerr}
\eeq
(Eq. (4.8) in \cite{Teukolsky-1973})
where $K \equiv (r^2+a^2) \omega - am$. The angular separation constant $\Lambda = A_{lm} - 2ma\omega + a^2 \omega^2$ is determined from imposing physical boundary conditions at the poles ($\theta=0, \pi$) on the angular equation
\beq
\frac{d}{dz} \left[ (1-z^2) \frac{d S}{d z} \right] + \left[ a^2 \omega^2 z^2 - 2 a m \omega s z + s + A_{lm} - \frac{(m+sz)^2}{1-z^2} \right] S = 0, \quad \text{where}\; z= \cos \theta ,
\eeq
whose solutions are oblate spin-weighted spheroidal harmonics $S = {}_sS_{lm}(a \omega; z)$ \cite{Abramowitz-Stegun}.  An expansion of the angular eigenvalue in powers of $a \omega$ is given in \cite{Seidel-1989, Berti-Cardoso-Casals}.

Let us make the substitution $R(r) = r^{-1} \Delta^{-s} u(r)$ to write the radial equation (\ref{rad-eq-kerr}) as
\beq
\frac{d^2 u}{d r_\ast^2} - \frac{2s(r-M)}{r^2} \frac{d u}{d r_\ast} + \left[ \Omega^2 - \frac{2 i s(r-M)\Omega}{r^2}  - f \left( \frac{\Lambda}{r^2} + \frac{f^\prime}{r} + \frac{2Ms}{r^3} - \frac{4 i \omega s}{r} \right)  \right] u = 0    \label{kerr-rad-eq2}
\eeq
where 
\beq 
f(r) = \Delta / r^2 = 1 - 2M / r + a^2 / r^2    \label{f-def}
\eeq
and 
\beq
\Omega(r) = K(r) / r^2 = (1 + a^2 / r^2) \omega - a m / r^2  . \label{Omega-def}
\eeq
Here we have defined a tortoise coordinate $r_\ast$ via
\beq
\frac{dr_\ast}{dr} = f^{-1}.
\eeq
Note that this definition differs from, e.g., Teukolsky \cite{Teukolsky-1973}, who defines instead $\tfrac{dr_\ast}{dr} = (r^2+a^2) / \Delta$.

Quasinormal modes are the complex-frequency modes that are purely ingoing at the (outer) horizon ($r=r_{h+}$), and purely outgoing at spatial infinity, satisfying boundary conditions
\beq
u(r) \sim \left\{ \begin{array}{ll} \exp( - i [\omega - a m / r_{h+}^2] r_\ast ),  \quad \quad &  r_\ast \rightarrow - \infty , \\ 
\exp( + i \omega r_\ast ), & r_\ast \rightarrow + \infty . \end{array} \right.   \label{qnm-bc}
\eeq

\section{QNM Expansion Method\label{sec:expansion-method}}
In this section we extend and develop the QNM expansion method first described in \cite{Dolan-Ottewill}. 
The method has three steps: (i) introduce an ansatz for the wavefunction, inspired by the null geodesics near the perpetual orbit, (ii) expand the wavefunction and frequency in inverse powers of $m$ (or $l$), (iii) impose a continuity condition at the perpetual orbit radius (i.e. $r = \hat{r}_\pm$ or $r = \hat{r}_\circ$) to recover the expansion coefficients. 

\subsection{Equatorial modes of scalar field\label{sec:eq-scalar}}
Let us illustrate the method by considering first the maximally corotating modes, $m = l$, of the scalar wave equation ($s = 0$).  We seek solutions to (\ref{kerr-rad-eq2}) that satisfy the QNM boundary conditions (\ref{qnm-bc}). 
Step (i) is to propose an ansatz of the form
\beq
u(r) = \exp \left( i \int  \beta(r) d r_\ast \right) v(r)  .  \label{ansatz}
\eeq
We demand that $\beta(r)$ has three properties. 
First, $\beta(r)$ should have a root at the circular orbit $r=\rcp$, so that $\beta$ changes sign here, and the solution passes smoothly from outgoing behaviour at infinity to ingoing behaviour at the horizon. 
Second, $\beta^2$ should be such that, upon insertion into Eq. (\ref{kerr-rad-eq2}), the resulting equation has an overall factor of $f$. That is, 
\beq
- \beta^2 + \Omega^2 \propto f ,  \label{second-condition}
\eeq
Third, $\beta$ should ensure that the QNM boundary conditions are satisfied, provided $v$ is regular at both horizon and infinity. 

The third property suggests that $\beta$ has an overall factor of $\Omega(r)$ (defined in Eq. (\ref{Omega-def})), and the first property suggests that we take inspiration from the factorized form of the orbital equation (\ref{equatorial-orbit-crit}) to write 
\beq
\beta(r) = \Omega(r) \left( 1 - \frac{a (b - a)}{r^2} \right)^{-1} \left(1 - \frac{\rcp}{r} \right) \left( 1 + \frac{2 \rcp}{ r} \right)^{1/2}
\eeq
where $b = b_+$, defined in (\ref{bcrit}). 
It is straightforward to confirm that the second property (\ref{second-condition}) is satisfied,
\beq
- \beta^2 + \Omega^2 = f(r) \, \Omega^2 \frac{(b-a)^2}{r^2} \left[1 - \frac{a(b-a)}{r^2} \right]^2 .
\eeq
and that the QNM boundary conditions are satisfied if $v(r)$ is regular at the horizon and infinity.  
Upon substitution of the ansatz (\ref{ansatz}) into (\ref{kerr-rad-eq2}), and after dividing through by $f$, we reach the new radial equation
\beq
\frac{d}{dr} \left( f  \frac{dv}{dr} \right) + 2 i \beta \frac{dv}{dr} + \left[ \Omega^2 \frac{(b-a)^2}{r^2} \left(1 - \frac{a (b-a)}{r^2}  \right)^{-2} + i \beta^\prime - \frac{\Lambda}{r^2} - \frac{f^\prime}{r} \right] v = 0 .
\eeq

Step (ii) is to expand the wavefunction, frequency and angular eigenvalue in inverse powers of $m=l$, in the following manner,
\begin{eqnarray}
b \, \omega &=&  \varpi_{-1} m + \varpi_{0} + \varpi_{1} m^{-1} + \ldots \label{omexpansion} \\
v(r) &=& \exp\left( S_0(r) + m^{-1} S_1(r) + \ldots \right) \label{vexpansion} \\
\Lambda &=& \Lambda_{-2} m^2 + \Lambda_{-1} m + \Lambda_{0} + \ldots \label{Lamexpansion}
\end{eqnarray}
Here, $\{\varpi_{-1}, \varpi_{0}, \ldots \}$ and $\{\Lambda_{-2}, \Lambda_{-1}, \ldots \}$ are coefficients to be determined, and $\{ S_0(r), S_1(r), \ldots \}$ are regular functions of $r$. 
Inserting (\ref{omexpansion}, \ref{vexpansion}, \ref{Lamexpansion}) into (\ref{kerr-rad-eq2}) and grouping together like powers of $m$ leads to a system of equations,
\begin{eqnarray}
\mathcal{O}(m^2) \, : \; && \frac{(b-a)^2}{b^2} \left[ \left(1 + \frac{a^2}{r^2}  \right) \varpi_{-1} - \frac{a b}{r^2}  \right]^2 \left(1 - \frac{a(b-a)}{r^2} \right)^{-2} - \Lambda_{-2} = 0  \label{expansion-ord0} \\
\mathcal{O}(m^1) \, : \; && 2 i \beta_0 \frac{d S_0}{d r} + i \beta_0^\prime - \frac{\Lambda_{-1}}{r^2} + 2\left(1 + \frac{a^2}{r^2} \right) \varpi_{0} \frac{(b-a)^2}{b^2 r^2} = 0  \label{expansion-ord1} \\
\mathcal{O}(m^0) \, : \; && \ldots
\end{eqnarray}
Here $\beta_0(r) = (\varpi_{-1} / b) (1 - a(b-a)/r^2)^{-1} (1-\rcp/r) ( 1 + 2 \rcp / r )^{1/2}  $ and $b = b_+$ given in Eq.~(\ref{bcrit}). 
A major difference with the Schwarzschild case is that the angular expansion coefficients $\Lambda_{k}$ is also dependent on frequency, through $a \omega$. This challenge is not insurmountable. First we note that $\Lambda_{-2} = (1 - a \varpi_{-1} / b)^2$. It is trivial to then show that the choice $\varpi_{-1} = 1$ satisfies Eq.~(\ref{expansion-ord0}). To find higher-order coefficients,  we may use the series expansion of the eigenvalue given in \cite{Seidel-1989, Berti-Cardoso-Casals}. We find
\begin{eqnarray}
\Lambda_{-2} &=& (1 - x)^2 \\
\Lambda_{-1} &=& \mathcal{S}_1 - 2 x (1 - x) \varpi_{0}  \label{lambda-1} \\
\Lambda_{0} &=& - 2s^2 x \mathcal{S}_2 + \frac{3x^2}{4} \mathcal{S}_3 + x\frac{d \mathcal{S}_1}{dx} \varpi_{0}
\end{eqnarray}
where $x = a / b$ and 
\begin{eqnarray}
\mathcal{S}_1 = 1 - \tfrac{1}{2} x^2 - \tfrac{1}{8} x^4 - \tfrac{1}{16} x^6 + \mathcal{O}(x^8) &\doteq& \left( 1 - x^2 \right)^{1/2} \\
\mathcal{S}_2 = 1 + x + x^2 + x^3 + x^4 + x^5 + \mathcal{O}(x^6) &\doteq& (1 - x)^{-1} \\
\mathcal{S}_3 = 1 + x^2 + x^4 + \mathcal{O}(x^6) &\doteq& (1 - x^2)^{-1}
\end{eqnarray}
Though Seidel \cite{Seidel-1989} provides an expansion to sixth order in $a \omega$, this provides only the first few terms of a (presumably) infinite series. Above, we used intuition to `guess' a closed form from the first few terms of the series expansion (the results to the right of the $\doteq$ symbol). Mathematica can be used to obtain further terms in the expansion of the spheroidal eigenvalue in the scalar case ($s=0$); the higher terms are consistent with the assumption.  In Sec.~\ref{sec-validation} we test the resulting frequencies against numerical results and find the expected agreement.

Next we demand that the radial function $S_0(r)$, which features in the expansion of the wavefunction (\ref{vexpansion}), is continuous at $r=\rcp$ (where $\beta(\rcp) = 0$). After inserting result (\ref{lambda-1}) into (\ref{expansion-ord1}) and evaluating at $r=\rcp$ and rearranging, we obtain the coefficient
\beq
\varpi_{0} =  \frac{1-2x}{2\sqrt{1 - x^2}}  (1 - i)  \label{omco0}
\eeq
The function $S_0$ may then be determined by substituting (\ref{omco0}) back into (\ref{expansion-ord1}) and rearranging to find $S_0^\prime$, then integrating. 

In principle, we may continue in this fashion to determine the higher-order coefficients $\varpi_1, \varpi_2, \ldots$ and phase functions $S_1(r), S_2(r), \ldots$, with the help of a symbolic algebra package. In practice it is a challenge to continue the expansion beyond $\mathcal{O}(m^{-2})$, due to the frequency dependence of the angular eigenvalue. Results to this order are given in Eq.~(\ref{corotating-expansion}).

To look for higher modes ($n > 0$) one must first modify the ansatz for the wavefunction (\ref{vexpansion}), as shown in \cite{Dolan-Ottewill}. This technique follows through to the axisymmetric case without additional difficulties, although the calculation is not pursued here.

\subsection{Equatorial modes of higher spin}
Let us now show how one may generalise the previous analysis to treat fields of higher spin ($s \neq 0$). 
As in Sec. \ref{sec:eq-scalar}, we start with Eq. (\ref{kerr-rad-eq2}) and once again propose an ansatz of the form (\ref{ansatz}). As in Sec. \ref{sec:eq-scalar}, we demand that $\beta$ obeys three conditions, but now the second condition (\ref{second-condition}) becomes
\beq
- \left( \beta  + \frac{i s (r-M)}{r^2} \right)^2 + \left(\Omega - \frac{is (r-M)}{r^2}\right)^2 \propto f ,  \label{second-condition-spin}
\eeq
The conclusion is that
\beq
\beta + \frac{is(r-M)}{r^2} = \left( \Omega - \frac{is (r-M)}{r^2} \right) \left(1 - \frac{a(b-a)}{r^2} \right)^{-1} \left( 1 - \frac{\rcp}{r} \right) \left( 1 + \frac{2 \rcp}{r} \right)^{1/2} .
\eeq
It is straightforward to verify that $\beta$ satisfies the QNM boundary conditions at the horizon and at infinity.

The method proceeds through step (ii) and (iii) as in the scalar case. The method was automated using a symbolic algebra package. We now present the key results for the equatorial modes.

\subsection{Equatorial Modes: Key Results}
Applying the method described above, we find that the fundamental ($n=0$) maximally \emph{co-rotating} ($m=l$) frequency $\omega^{(m=l)}_{n=0}$ has the following `eikonal' expansion :
\beq
b_+ \, \omega^{(m=l)}_{l,n=0} = m + \frac{(1-2x)}{2 \sqrt{1-x^2}} (1 - i) +  \frac{(1-2x)}{216 m (1-x^2)^2} \left[ (7+44x+127x^2) - 72 s^2 (1+x)^2 \right] +
\mathcal{O}(m^{-2})  \label{corotating-expansion}
\eeq
where here $x = x_+ \equiv a / b_{+}$
and the critical impact parameter $b_+$ was defined in (\ref{bcrit}). We established in Eq. (\ref{kepler-freq}) that $1/b_+$ is equal to the Kepler orbital frequency $\omega_+$ for null rays in corotating circular orbit. 
Furthermore, the leading-order imaginary component is equal to one-half of the Lyapunov exponent given in (\ref{lyapunov-eq}) (and Refs. \cite{Mashhoon, Lyapunov1}). 
%\beq
%\frac{(1-2x)}{2 \sqrt{1-x^2}} = \frac{(12M)^{1/2}[(r_c^+)^2 - 2M r_c^+ + a^2]}{(r_c^+)^{3/2} (r_c^+ - M)}
%\eeq

To leading order, our result is consistent with the outcome of the geodesic analysis of Mashhoon \cite{Mashhoon}, Cardoso \emph{et al.} \cite{Lyapunov1} and Hod (\cite{Hod}, Eq. 3 and 4), and with the wave-equation analysis of Hod in the extremal regime. Note however that result (\ref{corotating-expansion}) goes two steps further. Firstly, it includes a spin-independent correction to the real part of frequency at order $\mathcal{O}(m^0)$, which has not been previously obtained. Secondly, it also includes the spin-dependent correction at order $\mathcal{O}(m^{-1})$ for the first time. In Sec. \ref{sec-validation} we compare result (\ref{corotating-expansion}) against numerically-determined frequencies, to verify that both new terms are correct. %to order $\mathcal{O}(m^{-2})$.

The expansion of the fundamental ($n=0$) maximally \emph{counter-rotating} ($m=-l$) frequency $\omega_{n=0}^{(m=-l)}$ is 
\beq
b_- \, \omega^{(m=-l)}_{l,n=0} = m - \frac{(1-2x)}{2 \sqrt{1-x^2}} (1 - i) +  \frac{(1-2x)}{216 m (1-x^2)^2} \left[ (7+44x+127x^2) - 72 s^2 (1+x)^2 \right] +
O(m^{-2})  \label{counterrotating-expansion}
\eeq
where here $x = x_- \equiv a / b_-$. Note that the counter-rotating impact parameter $b_-$ given by (\ref{bcrit}) is \emph{negative}, as is $x$ (for positive $a$). Result (\ref{counterrotating-expansion}) may be obtained from result (\ref{corotating-expansion}) via the simultaneous replacements $b_+ \rightarrow b_-$, $x_+ \rightarrow x_-$, $m \rightarrow -m$ and $\omega_{n=0}^{(m=l)} \rightarrow -\omega_{n=0}^{(m=-l)}$.

The QNM spectrum has the following symmetry property:
\beq
\omega_{lmn} = - \omega_{l, -m, n}^\ast .  \label{qnm-symmetry}
\eeq
Therefore frequencies with negative real part may be obtained by applying (\ref{qnm-symmetry}) to Eqs.~(\ref{counterrotating-expansion}) and (\ref{corotating-expansion}).

\subsection{QNMs of polar modes ($m=0$)}
In Sec. \ref{subsec-polar-geodesics} we considered `polar' geodesics, which have zero azimuthal angular momentum $L_z = 0$ and pass through the north and south poles $\theta = 0, \pi$. Here we apply the expansion method to the Teukolsky equation to find the corresponding $m=0$ QNM frequencies. 

For brevity, let us consider scalar waves ($s=0$) here, as it turns out that the leading order terms (at $l^1$ and $l^0$) are not spin-dependent. The appropriate ansatz for the polar modes is again of the form (\ref{ansatz}), with condition (ii) (Eq. (\ref{second-condition})) suggesting the choice
\beq
\beta^2 = \Omega^2 - f(r) \frac{\omega^2 b_\circ^2}{r^2} = \frac{ \omega^2 R(r) }{ r^4 }, \label{alpha-polar}
\eeq
where $b_\circ$ is the critical impact parameter (\ref{bcrit-polar}) and $R(r)$ is the quartic defined in Eq.~(\ref{R-defn}), with factorization given by Eq.~(\ref{R-factorization}). 

Inserting the ansatz (\ref{alpha-polar}) into (\ref{kerr-rad-eq2}) leads to
\beq
\left( f v' \right)^\prime + 2 i \beta(r) v'  + \left[ i \beta' + \frac{\omega^2 b_c^2 - \Lambda}{r^2} - \frac{f'}{r^2} \right] v  = 0 .  \label{kerr-eq-v}
\eeq
Next we expand in inverse powers of $L = l+1/2$, 
\begin{eqnarray}
b_\circ \omega &=&  \varpi_{-1} L + \varpi_0 +  \varpi_1 L^{-1} + \ldots \\
v(r) &=& \exp \left( S_0(r) + S_1(r) L^{-1} + \ldots \right) \\
\Lambda &=& \Lambda_{-2} L^2 + \Lambda_{-1} L + \Lambda_0 + \ldots
\end{eqnarray}
Now group terms in (\ref{kerr-eq-v}) order-by-order in $L$,
\begin{eqnarray}
\frac{\varpi_{-1}^2 - \Lambda_{-2}}{r^2} &=& 0 \label{polar-expansion0}  \\
\frac{2 i \varpi_{-1} \left( r^2 + 2r \rco - a^2 d^2 / \rco^2 \right)}{r^2}  (r-\rco)  S^\prime_0(r)  +  \frac{2 \varpi_{-1} \varpi_0 }{r^2} + i \beta_0^\prime - \frac{\Lambda_{-1}}{r^2} &=& 0 \label{polar-expansion1}  \\
\ldots &=& 0
\end{eqnarray}
%It is immediately clear from Eq.~(\ref{polar-expansion0}) that $\varpi_{-1} = \pm \sqrt{ \Lambda_{-2} }$.  However, 
Once again, there remains an obstacle to progress: the frequency-dependence of the angular eigenvalue $\Lambda$. 
The angular eigenvalue of the scalar field ($s=0$) for the polar mode $\Lambda_{l,m=0}$ has the following expansion:
\beq
\Lambda_{l,m=0} = L^2 + \left( \frac{c^2}{2} - \frac{1}{4} \right) + \frac{c^4 - 4c^2}{32 L^2} + \frac{5}{64} \frac{c^4 - 8c^2}{L^4} + \frac{5c^8 - 160 c^6 + 2256 c^4 - 1024 c^2}{8192 L^6} + c^2 \mathcal{O}\left( \frac{c^8}{L^8} \right) ,  \label{Lambda-expansion}
\eeq
where $c = a \omega$ and $L=l+1/2$. This result is given in Ref.~\cite{Rokhlin-Xiao}, Theorem 10. For higher spin fields $s \neq 0$ we may use the expansions given in \cite{Seidel-1989, Berti-Cardoso-Casals}: 
\begin{eqnarray}
\Lambda &=& L^2 \left[ 1 + \frac{1}{2} \frac{c^2}{L^2} + \frac{1}{32} \frac{c^4}{L^4} + 0 + \mathcal{O}\left( \frac{c^8}{L^8} \right) \right] \nonumber \\
 && \; + \left[ -s(s+1) - 1/4 + \left(-\frac{1}{8} + s^2 \right) \frac{c^2}{L^2} + \left(\frac{5}{64} - \frac{3s^2}{8} \right) \frac{c^4}{L^4} + \left( \frac{-5}{256} + \frac{5 s^2}{32} \right) \frac{c^6}{L^6} + \mathcal{O}\left(\frac{c^8}{L^8} \right) \right]
\end{eqnarray}
This expansion allows us to write $\Lambda_{-2}$ and $\Lambda_{-1}$ as power series in $\alpha \varpi_{-1}$, where
\begin{eqnarray}
\Lambda_{-2} &=& 1 + \frac{1}{2} (\alpha \varpi_{-1})^2 + \frac{1}{32} (\alpha \varpi_{-1})^4 + 0 + \frac{5}{8192}(\alpha \varpi_{-1})^8 + \mathcal{O} ( (\alpha \varpi_{-1})^{10} ) , \\
\Lambda_{-1} &=& (\alpha \varpi_{-1}) (\alpha \varpi_{0}) \left[ 1 + \frac{1}{8} (\alpha \varpi_{-1})^2 + 0 + \frac{5}{1024} (\alpha \varpi_{-1})^6  + \mathcal{O} ( (\alpha \varpi_{-1})^8 ) \right] ,
\end{eqnarray}
and
\beq
\alpha \equiv a / b_\circ .
\eeq
Hence to find $\varpi_{-1}$ we must solve a non-linear equation. This is straightforward to do iteratively. We may obtain a converging sequence of estimates $\{ \varpi_{-1}^{[0]} = 1, \varpi_{-1}^{[1]} , \varpi_{-1}^{(2)}, \ldots   \}$ using
\beq
\varpi_{-1}^{[k+1]} = \left( 1 +  \frac{1}{2} (\alpha \varpi_{-1}^{[k]})^2 + \frac{1}{32} (\alpha \varpi_{-1}^{[k]} )^4 + 0 + \frac{5}{8192} (\alpha \varpi_{-1}^{[k]})^{8}  + \mathcal{O}( (\alpha \varpi_{-1})^{10} ) \right)^{1/2} \label{polar-om-1}
\eeq
To find $\varpi_0$ we impose a continuity condition on $S_0(r)$ at $r = \rco$, and solve to obtain
\beq
\varpi_{0} = \frac{-i (3\rco^2 - a^2 \dco^2 / \rco^2)^{1/2}}{2 b_\circ} \left[  1 -  \frac{\alpha^2}{2} \left( 1 + \frac{1}{8}(\alpha \varpi_{-1})^2 + 0 + \frac{5}{1024} (\alpha \varpi_{-1})^6 + \ldots \right) \right]^{-1}  \label{polar-om0}
\eeq
Note that $\varpi_{-1}$ is purely real and $\varpi_{0}$ is purely imaginary. 

It is unfortunate that closed-form results are not so easily obtained in the polar ($m=0$) case. However, we may compare Eq.~(\ref{polar-om-1}) with Eq.~(\ref{polar-orbfreq}), and Eq.~(\ref{polar-om0}) with Eq.~(\ref{polar-lyapunov}), by evaluating numerically for a given rotation parameter $a$. It turns out that the numerical values are in precise agreement. For example, for $a = 0.8M$ we have $\rco \approx 2.67062M$, $\bco \approx 4.98488M$ and $\dco \approx 4.92027M$. Iteration of Eq.~(\ref{polar-om-1}) leads to $\varpi_{-1}/\bco \approx 0.20191308580895 \, 795 M^{-1}$, which should be compared with $\omega_\circ \approx 0.20191308580895 \, 903 M^{-1}$ from Eq. (\ref{polar-orbfreq}), i.e. agreement to 14 significant figures. Equation (\ref{polar-om0}) leads to $\varpi_{0} / \bco = -0.0893785765669 \, 3478 i M^{-1}$, which should be compared with $-i \lambda_\circ / 2 = -0.0893785765669 \, 4054 i M^{-1}$ from Eq. (\ref{polar-lyapunov}), i.e. agreement to 12 significant figures. 

We conclude that the polar ($m=0$) modes have the following frequency expansion:  
\begin{eqnarray}
\omega_{l,n=0}^{(m=0)} &=& \omega_\circ (l+1/2) - i \lambda_\circ / 2 + \mathcal{O}(m^{-1})  \nonumber \\
 &=&   (\varpi_{-1} / b_\circ ) (l+1/2) + (\varpi_{0} / b_\circ) (2n+1) + \mathcal{O}(m^{-1})   \label{polar-expansion}
\end{eqnarray}
where $\{ \omega_\circ, \lambda_\circ \}$ are given in Eq.~(\ref{polar-orbfreq}) and (\ref{polar-lyapunov}), and $\{ \varpi_{-1}, \varpi_{0} \}$ are given in Eq.~(\ref{polar-om-1}) and (\ref{polar-om0}). In Sec.~\ref{sec-validation} we compare this estimate against numerically-determined frequencies.

%One benefit of the expansion in powers of $L$ is that (at least for $m=0$), $\omega_n$ is purely real when $n$ is even and purely imaginary when $n$ is odd.

% Flammer. However, I think Flammer's coefficient $l_8$ is incorrect, because its leading order behaviour in $L$ goes as $L^{-4}$ (too high).  All other references giving a series expansion in small $a \omega$ merely reproduce Flammer (e.g. Abramowitz and Stegun, Breuer et al, etc). Seidel only goes up to 6th order.

%\section{Results}

%\subsection{Frequency Expansions}

\section{Validation\label{sec-validation}}
In this section we test the key results (\ref{corotating-expansion}), (\ref{counterrotating-expansion}) and (\ref{polar-expansion}) by comparing with numerically-determined frequencies. A fast and accurate numerical method for determining QNM frequencies for the Kerr black hole was introduced by Leaver \cite{Leaver-1985} many years ago. We have implemented our own version of the continued-fraction algorithm that Leaver described. Other implementations are available in the public domain \cite{Berti-webpage, Berti-Cardoso-Starinets}.

\subsubsection{Equatorial Modes}
%New approximations for QNM frequencies of maximally co-rotating ($m=l$) and counter-rotating ($m = -l$) modes were given in Eq.~(\ref{corotating-expansion}) and Eq.~(\ref{counterrotating-expansion}). The new expressions improve upon previous approximations \cite{Mashhoon, Lyapunov1, Hod} in two regards, giving: (i) a spin-independent correction to the real part of the frequency at order $m^{0}$, (ii) a spin-dependent correction to the real part at order $m^{-1}$. The aim of this section is to check these corrections through comparison with numerically-determined frequencies.

Table \ref{table:eq-scalar} shows the fundamental ($n=0$) frequencies for scalar-field ($s=0$) modes with $|m| = l = 2,4,6,8,10$ at $a = 0.8M$. For each $l$, the upper row gives the numerically-determined frequency. The lower rows give the approximations at 0th ($m^{1}$), 1st ($m^0$) and 2nd ($m^1$) orders, obtained from Eq.~(\ref{corotating-expansion}) and (\ref{counterrotating-expansion}). Previous approximations \cite{Mashhoon, Lyapunov1, Hod} only supplied the real part to order $m^{1}$ (0th), and the imaginary part to order $m^{0}$ (1st). The table shows that including the higher-order corrections to the real part improves the accuracy of the estimate substantially. In fact, even at relatively low $l$, Eq.~(\ref{corotating-expansion}) and (\ref{counterrotating-expansion}) are surprisingly accurate estimates of the scalar-field frequencies. For example, at $l=2$ the estimate of the real part is accurate to $0.3\%$ (co-rotating) and $0.06\%$ (counter-rotating). The estimate of the imaginary part is accurate to $1.1\%$ (co-rotating) and $0.6\%$ (counter-rotating). 

Table \ref{table:eq-gravitational} shows the fundamental ($n=0$) frequencies for the gravitational field ($|s| = 2$) at $a = 0.8M$. Again, it is clear that including the higher-order corrections improves the accuracy of the estimate. However, the magnitude of the error in the estimate is significantly greater than for the scalar field, implying that field spin has a non-negligible effect outside the eikonal regime (i.e. at small or moderate $l$). For example, the real part of the $l=2$ estimate is in error by $10.4 \%$ (co-rotating) and $2.2\%$ (counter-rotating), and by $6.6\%$ and $5.9\%$ for the imaginary part. In the low-$l$ regime the expansion is substantially less accurate than the WKB method \cite{Seidel-Iyer, Kokkotas}.% (though less opaque).

\begin{table}
\begin{tabular}{r r | r r | r r}
\hline
\hline
 & & Prograde & $(m=l)$ & Retrograde & $(m = -l)$ \\
$|m|$ & & Re($M \omega$)  & Im($M \omega$) & Re($M \omega$) & Im($M \omega$) \\
\hline
$|m| = 2$, & exact  & 0.70682338 & -0.08152026 & 0.39573382 & -0.09428527 \\
& 2nd & 0.70892744 & &  0.39550921 & \\
& 1st & 0.69841410 & -0.08061490 & 0.39393550 & -0.09374776 \\
& 0th & 0.61779920 & & 0.30018774 & \\
\hline
$|m| = 4$, & exact  & 1.32089554 & -0.08088308 & 0.69498056 & -0.09393068 \\
& 2nd & 1.32146997 & & 0.69491010 & \\
& 1st & 1.31621330 &  -0.08061490 & 0.69412324 & -0.09374776 \\
& 0th & 1.23559839 & & 0.60037549 & \\
\hline
$|m| = 6$, & exact  & 1.93725163 & -0.08074051 & 0.99486912 & -0.09383851 \\
& 2nd & 1.93751694 & & 0.99483556 & \\
& 1st & 1.93401249 &  -0.08061490 & 0.99431099 & -0.09374776 \\
& 0th & 1.85339759 & & 0.90056323 & \\
\hline
$|m| = 8$, & exact  & 2.55428739 & -0.08068741 & 1.29491167 & -0.09380178 \\
& 2nd & 2.55444002 & & 1.29489216 & \\
& 1st & 2.55181169 &  -0.08061490 & 1.29449873 & -0.09374776 \\
& 0th & 2.47119678 & & 1.20075097 & \\
\hline
$|m| = 10$, & exact  & 3.17161440 & -0.08066204 & 1.59501395 & -0.09378355 \\
& 2nd & 3.17171355 & & 1.59500121 & \\
& 1st & 3.16961088 &  -0.08061490 & 1.59468647 & -0.09374776 \\
& 0th & 3.08899598 & & 1.50093872 & \\
\hline
\hline
\hline 
\end{tabular}
\caption{\emph{Equatorial Modes: Scalar Field}. For each $|m|$, the top row is the numerically-determined QNM frequency of the fundamental mode, for $m=l$ (left) and $m = -l$ (right). The lower rows give the estimates from Eq.~(\ref{corotating-expansion}) and (\ref{counterrotating-expansion}) at orders $m^1$ (0th), $m^0$ (1st) and $m^{-1}$.}
\label{table:eq-scalar}
\end{table}

\begin{table}
\begin{tabular}{r r | r r | r r}
\hline
\hline
 & & Prograde & $(m=l)$ & Retrograde & $(m = -l)$ \\
$|m|$ & & Re($M \omega$)  & Im($M \omega$) & Re($M \omega$) & Im($M \omega$) \\
\hline
$|m| = 2$, & exact  & 0.58601697 & -0.07562955 & 0.30331342 & -0.08851224 \\
& 2nd & 0.52518088 & & 0.29659659 & \\
& 1st & 0.69841410 & -0.08061490 & 0.39393550 & -0.09374776 \\
& 0th & 0.61779920 & & 0.30018774 & \\
\hline
$|m| = 4$, & exact  & 1.24754701 & -0.07812549 & 0.64854056 & -0.09258611 \\
& 2nd & 1.22959668 & & 0.64545379 & \\
& 1st & 1.31621330 & -0.08061490 & 0.69412324 & -0.09374776 \\
& 0th & 1.23559839 & & 0.60037549 & \\
\hline
$|m| = 6$, & exact  & 1.88474955 & -0.07927839 & 0.96349484 & -0.09323669 \\
& 2nd & 1.87626808 & & 0.96186468 & \\
& 1st & 1.93401249 & -0.08061490 & 0.99431098 & -0.09374776 \\
& 0th & 1.85339759 & & 0.90056323 & \\
\hline
$|m| = 8$, & exact  & 2.51341639 & -0.07979031 & 1.27116161 & -0.09345939 \\
& 2nd & 2.50850338 & & 1.27016400 & \\
& 1st & 2.55181169 & -0.08061490 & 1.29449873 &  -0.09374776\\
& 0th & 2.47119678 & & 1.20075097 & \\
\hline
$|m| = 10$, & exact  & 3.13816080 & -0.08005731 & 1.57589011 & -0.09356231 \\
& 2nd & 3.13496424 & & 1.57521869 & \\
& 1st & 3.16961088 & -0.08061490 & 1.59468647 & -0.09374776 \\
& 0th & 3.08899598 & & 1.50093872 & \\
\hline
\hline
\hline 
\end{tabular}
\caption{\emph{Equatorial Modes: Gravitational Field}. As Table \ref{table:eq-scalar} but for the gravitational QNMs.}
\label{table:eq-gravitational}
\end{table}

\begin{figure}
 \includegraphics[width=8cm]{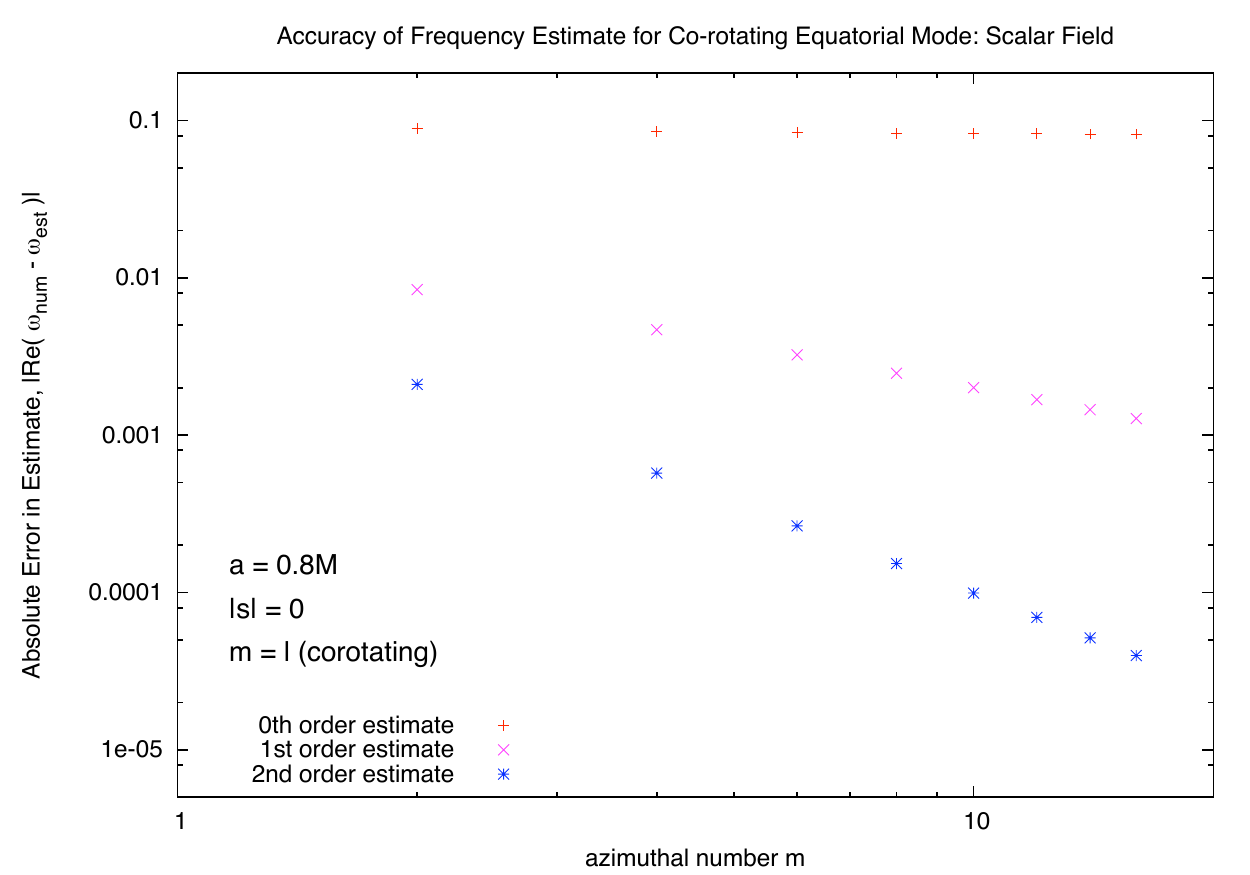}
 \includegraphics[width=8cm]{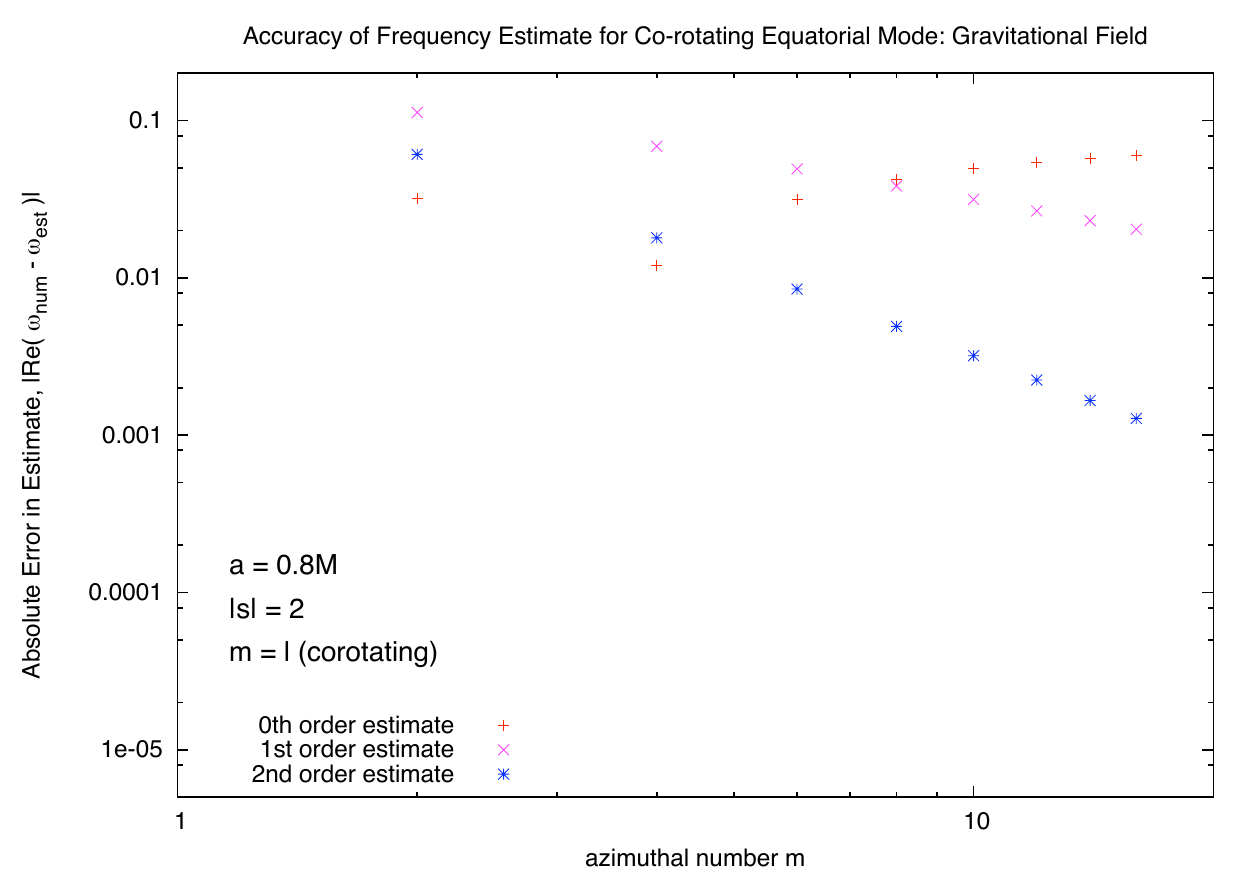}
 \includegraphics[width=8cm]{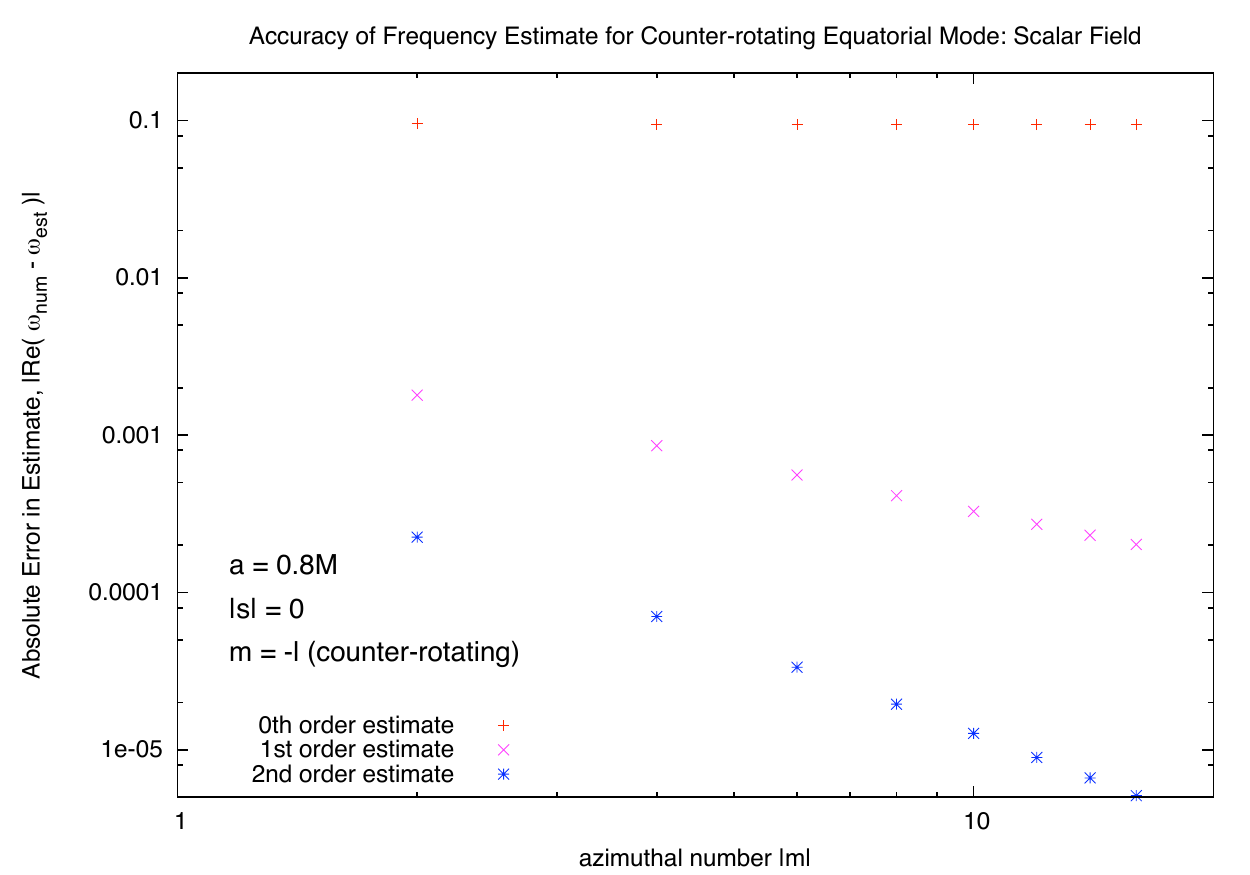}
 \includegraphics[width=8cm]{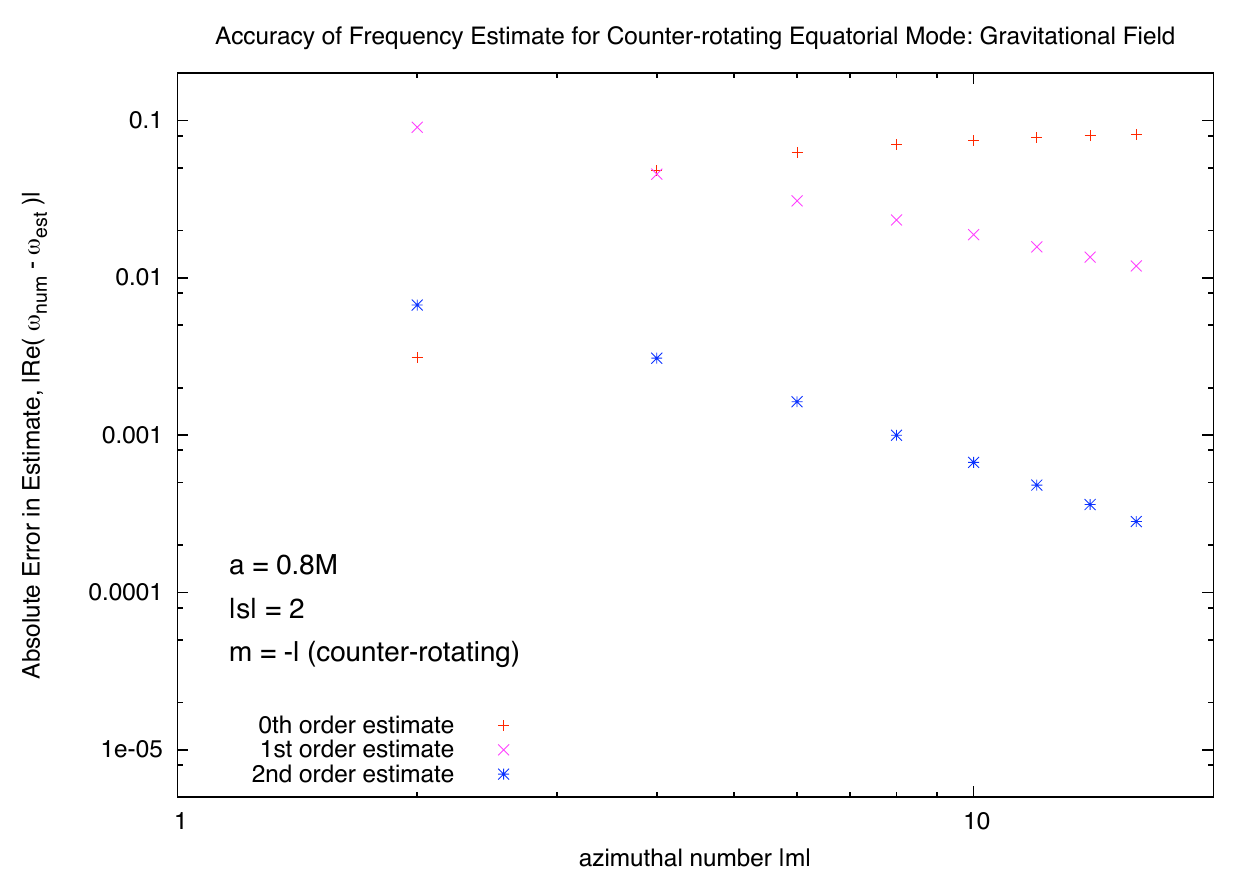}
 \includegraphics[width=8cm]{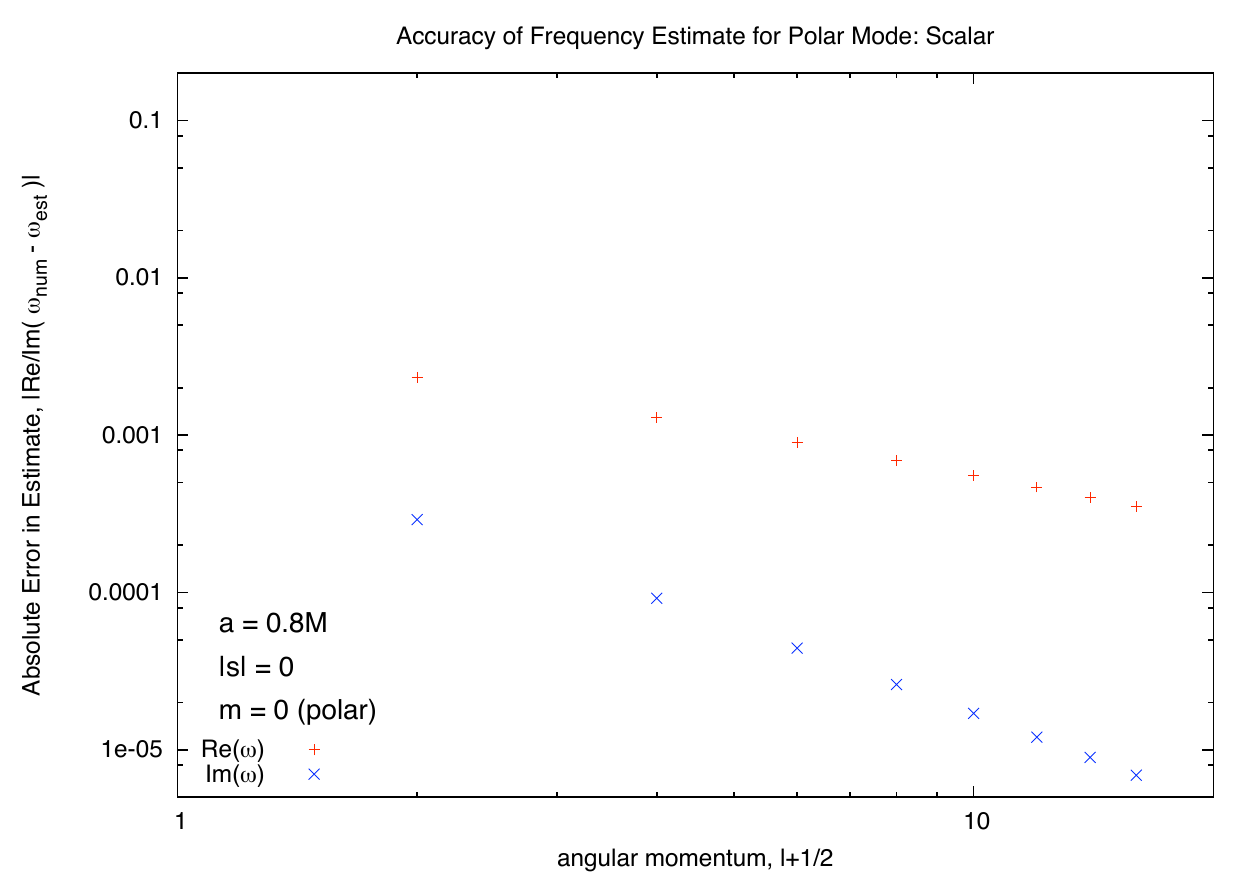}
 \includegraphics[width=8cm]{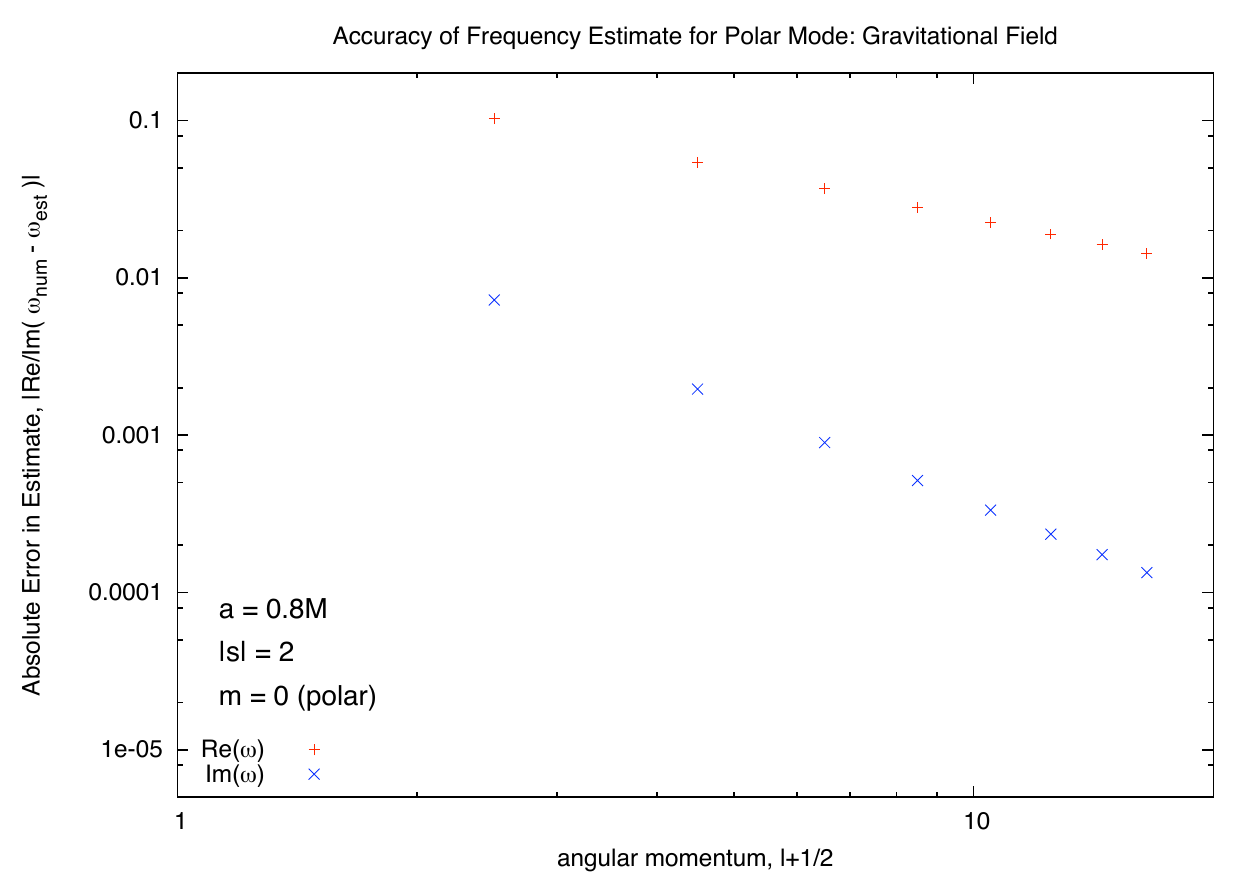}
 \caption{\emph{Accuracy of the QNM Frequency Estimates}. These log-scale plots show the difference between the numerically-determined QNM frequency and the estimates obtained in this paper, as function of angular momentum $l$. The left plots show the scalar field, and the right plots show the gravitational field. The upper ($m=l$) and middle ($m=-l$) plots show the difference in the real part of the frequency, for approximations (\ref{corotating-expansion}) and (\ref{counterrotating-expansion}) truncated at orders $l^{1}$ (0th order), $l^0$ (1st) and $l^{-1}$ (2nd). At large $l$, the slopes of the data sets on the log-log scale approach $0$, $-1$ and $-2$, as expected.  The lower plots show the difference for the real and imaginary parts of the polar ($m=0$) frequencies. The gradients tend to -1 [red, real] and -2 [blue, imaginary], implying that the approximation is accurate up to order $\mathcal{O}(L^{-1})$ for the real part and $\mathcal{O}(L^{-2})$ for the imaginary part.}
 \label{fig-accuracy}
\end{figure}

Figure \ref{fig-accuracy} shows the `error' (defined as difference between the estimate given in Sec.~\ref{sec:expansion-method} and the numerically-determined frequency) as a function of $m$, on a log-log scale. It provides strong evidence that the estimates given in (\ref{corotating-expansion}) and (\ref{counterrotating-expansion}) are indeed correct to the stated order in $m$. The upper plots show the error in the real part of frequency for the scalar (left) and gravitational (right) cases, for co-rotating orbits. The middle plots show the same for the counter-rotating orbits.  The data set marked ``0th" shows the error using only the order $m^1$ estimate. The data sets marked ``1st" and ``2nd" show the effect on error on including the order $m^0$ and order $m^{-1}$ corrections. The plots shows that the error scales as $m^0$ (0th), $m^{-1}$ (1st) and $m^{-2}$ (2nd) in the large-$m$ regime (this may be inferred by examining the `slope' of the respective data sets in the log-log plot, and confirming that it tends to $0$, $-1$ and $-2$ in the large-$l$ limit, for 0th, 1st and 2nd approximations). Comparing the left and right plots, it is clear that the absolute error is significantly greater for the gravitational field than for the scalar field. 

\subsubsection{Polar Modes}

Table \ref{table:polar} compares numerically-determined polar ($m=0$) frequencies for scalar and gravitational fields, with the approximation of Eq.~(\ref{polar-expansion}). The approximation is closer to the scalar QNM frequency than to the gravitational QNM frequency. In the lower plots of Fig.~\ref{fig-accuracy}, the error in the real and imaginary part of frequency is shown. It is clear from the gradients of the two data sets that the next-order correction to (\ref{polar-expansion}) is at $\mathcal{O}( l^{-1} )$ for the real part, and $\mathcal{O}(l^{-2})$ for the imaginary part.

\begin{table}
\begin{tabular}{r r | r r}
\hline
\hline
$l$ & & Re($M \omega$) & Im($M \omega$) \\
\hline
$l = 2$, & scalar & \quad  0.50712267 & \quad-0.08966931 \\
 & gravitational & 0.40191735 & -0.08215627 \\
 & estimate & 0.50478271 & -0.08937858 \\
\hline
$l = 4$, & scalar & 0.90990916 & -0.08947052 \\
 & gravitational & 0.85459105 & -0.08741415 \\
 & estimate & 0.90860889 & -0.08937858 \\
\hline
$l = 6$, & scalar & 1.31333515 & -0.08942290 \\
 & gravitational & 1.27557678 & -0.08848090 \\
 & estimate & 1.31243506 & -0.08937858 \\
\hline
$l = 8$, & scalar & 1.71694949 & -0.08940455 \\
 & gravitational & 1.68823375 & -0.08886365 \\
 & estimate & 1.71626123 & -0.08937858 \\
\hline
$l = 10$, & scalar & 2.12064455 & -0.08939561 \\
 & gravitational & 2.09746074 & -0.08904433 \\
 & estimate & 2.12008740 & -0.08937858 \\
\hline
\end{tabular}
\caption{\emph{Polar Modes}. The table compares the numerically-determined values of the `polar' ($m = 0$) modes (top two rows) with the lowest-order estimate (\ref{polar-expansion}). Note that at leading order the estimate is spin-independent.  }
\label{table:polar}
\end{table}

\section{Discussion and Conclusions\label{sec:conclusion}}

In this paper we have obtained new closed-form approximations for the fundamental ($n = 0$) Kerr QNM frequencies of the equatorial ($m = \pm l$) and polar ($m = 0$) modes of arbitrary spin, valid in the regime $l \gg |s|$; see Eq.~(\ref{corotating-expansion}), (\ref{counterrotating-expansion}) and (\ref{polar-expansion}). The result for the corotating modes improves upon previous approximations \cite{Mashhoon, Lyapunov1, Hod} in two regards, giving: (i) a spin-independent correction to the real part of the frequency at order $m^{0}$, (ii) a spin-dependent correction to the real part at order $m^{-1}$. The result for the polar modes is (we believe) substantially new. We demonstrated that the approximations fit the numerical data in the asymptotic regime. We have sought to improve understanding of the link between the asymptotic structure of the QNM spectrum and the properties of the unstable photon orbits (see also \cite{Goebel, Mashhoon, Lyapunov1, Hod}).

Let us conclude by suggesting some possible extensions to this work: (i) it is simple to extend the analysis to higher overtones ($n > 0$) following \cite{Dolan-Ottewill}; (ii) it should be straightforward to extend the analysis to the Kerr-Newman ($Q \ge 0 $) black hole (though here coupling between electromagnetic and gravitational perturbations may complicate the issue \cite{Kokkotas-1993, Kokkotas-Berti-2005}); (iii) here we have considered only the limiting cases of equatorial ($|m| = l$) and polar ($m=0$) modes here; extension to general modes ($m = \mu l$, $-1 \le \mu \le 1$) remains an open challenge. We expect the real and imaginary parts to again be linked to the orbital frequency and Lyapunov exponent on the relevant null orbit (i.e. the general orbits considered in \cite{Teo}), although we also expect some non-trivial correction at $\mathcal{O}(l^0)$, such as appears in (\ref{corotating-expansion}); (iv) The expansion method detailed here could also be applied to find the Regge poles of the Kerr black hole in the eikonal regime \cite{Decanini-Folacci-2010a, Decanini-Folacci-2010b}; (v) The expansion method can also be applied to compute QNM excitation factors \cite{Leaver-1986,Berti-Cardoso-2006}, which are needed to tackle the question `how much QNM ringing is excited by a given perturbation?' We hope to pursue this calculation in the near future.

\appendix

%\subsection{Equatorial Orbits}
%Two useful identities are
%\beq
%\rcpm = \pm \frac{b_\pm}{\sqrt{3}} \left( 1 - x^2 \right)^{1/2}, \quad \quad M = \frac{\rcpm}{3} \left( \frac{1+x}{1-x} \right)   \label{identities-eq}
%\eeq

\section{Period and Lyapunov Exponent for Polar Orbit\label{appendix:polar}}
Here we derive expressions for the orbital frequency and Lyapunov exponent for the polar orbit.
The period $T_\circ$ with respect to the coordinate time is
\beq
T_\circ = 2 \int_{-1}^{1} \frac{\dot{t}}{\dot{z}} \, dz  ,
\eeq
where, from Eq.~(\ref{t-eq}) and (\ref{theta-eq}), 
\beq
\dot{t} = \left. \frac{ \Sigma^2 }{ \rho^2 \, \Delta} \right|_{r = \rco}, \quad \quad \quad 
\dot{z} = \left. \rho^{-2}  (\dco^2 + a^2 z^2)^{1/2} (1 - z^2)^{1/2} \right|_{r = \rco}.
\eeq
and $\dco = \sqrt{b_\circ - a^2} \equiv \sqrt{\mathcal{Q}} / E$. 
The $\dot{t}$ expression can be simplified with the use of the identities
\beq
\rco = \dco \left( \frac{\rco^2 - a^2}{3 \rco^2 + a^2} \right)^{1/2}, \quad \quad M = \rco \left( \frac{\rco^2 + a^2}{3 \rco^2 - a^2} \right) ,  \label{identities-polar}
\eeq
to obtain
\beq
\dot{t} = \rho^{-2} \left( \dco^2 + a^2 z^2 \right)
\eeq
Hence the period is
\beq
T_\circ = 2 \int_{-1}^{1} \frac{\left(\dco^2 + a^2 z^2\right)^{1/2}}{\left(1 - z^2 \right)^{1/2}} dz
\eeq
which evaluates to result (\ref{polar-period}).

The Lyapunov exponent $\lambda_\circ$ can be found from taking an orbital average,
\beq
\lambda_\circ = \frac{2}{T_\circ} \int_{-1}^{1} \bar{\lambda}\, \left( \frac{\dot{t}}{\dot{z}} \right) dz .
\eeq
of the expression \cite{Lyapunov1},
\beq
\bar{\lambda} = \left. \dot{t}^{-1} \sqrt{\frac{1}{2} \frac{d^2 V_r}{d r^2} } \right|_{r=\rco}
\eeq
where $V_r \equiv \dot{r}^2 = \rho^{-4} (r - \rco)^2 ( r^2 + 2 r \rco - a^2 \dco^2 / \rco^2 )$.
Using $V_r^{\prime \prime} = 2 \rho^{-4} (3 \rco^2 - a^2 \dco^2 / \rco^2)$ gives
\beq
\lambda_\circ = \frac{2 (3 \rco^2 - a^2 \dco^2 / \rco^2)^{1/2}}{T_\circ} \int_{-1}^{1}   \frac{dz}{\left(\dco^2 + a^2 z^2\right)^{1/2} \left(1 - z^2 \right)^{1/2}} .
\eeq
which leads directly to result (\ref{polar-lyapunov}).

\begin{acknowledgments}
I am grateful to Adrian Ottewill for many helpful suggestions and encouragement.  
Thanks also to Leor Barack, Emanuele Berti, Vitor Cardoso, Bahram Mashhoon, Marc Casals, Luis Crispino and Leandro Oliveira for interesting discussions which influenced this work. I acknowledge financial support from the Engineering and Physical
Sciences Research Council (EPSRC) under grant no. EP/G049092/1.
\end{acknowledgments}

\bibliographystyle{apsrev}

%\bibliography{year1}

\end{document}